\documentclass[12pt]{article}
\usepackage{mathpazo}
\usepackage[T1]{fontenc}
\usepackage[utf8]{inputenc}
\usepackage{graphicx}
\usepackage{url}

\usepackage{amsmath}
\usepackage{amsthm}
\usepackage{amssymb}
\usepackage{mathptmx}
\usepackage[numbers,round]{natbib}
\renewcommand{\thetable}{\Roman{table}}

\makeatletter
\def\tagform@#1{\maketag@@@{[\ignorespaces#1\unskip\@@italiccorr]}}
\makeatother

\renewcommand{\baselinestretch}{1}

\newcommand{\dea}{Dale \emph{et al.}}

\newcommand\blfootnote[1]{%
  \begingroup
  \renewcommand\thefootnote{}\footnote{#1}%
  \addtocounter{footnote}{-1}%
  \endgroup
}

\vspace{-1in}

\begin{document}
\title{Lack of evidence for a substantial rate of templated mutagenesis in B cell diversification}
\author{
  Running title: Lack of evidence for templated mutagenesis\\[1em]
  Julia Fukuyama\footnote{Department of Statistics, Indiana University Bloomington},
    Branden J Olson\footnote{Program in Computational Biology, Fred Hutchinson Cancer Research Center and Department of Statistics, University of Washington}, and
    Frederick A Matsen IV$^\dag$\blfootnote{Address correspondence to Frederick Matsen at \texttt{matsen@fredhutch.org}. This research was supported by National Institutes of Health grants R01 GM113246, R01 AI120961, U19 AI117891, and R01 AI146028.
The research of Frederick Matsen was supported in part by a Faculty Scholar grant from the Howard Hughes Medical Institute and the Simons Foundation.}
}
\date{}
\maketitle

\newpage

\begin{abstract}
B cell receptor sequences diversify through mutations introduced by purpose-built cellular machinery.
A recent paper has concluded that a ``templated mutagenesis'' process is a major contributor to somatic hypermutation, and therefore immunoglobulin diversification, in mice and humans.
In this proposed process, mutations in the immunoglobulin locus are introduced by copying short segments from other immunoglobulin genes.
If true, this would overturn decades of research on B cell diversification, and would require a complete re-write of computational methods to analyze B cell data for these species.

In this paper, we re-evaluate the templated mutagenesis hypothesis.
By applying the original inferential method using potential donor templates absent from B cell genomes, we obtain estimates of the methods's false positive rates.
We find false positive rates of templated mutagenesis in murine and human immunoglobulin loci that are similar to or even higher than the original rate inferences, and by considering the bases used in substitution we find evidence that if templated mutagenesis occurs, it is at a low rate.
We also show that the statistically significant results in the original paper can easily result from a slight misspecification of the null model.
\end{abstract}

\section*{\centering {\small Key points}}
\small
\begin{itemize}
\item
A recent study proposes a new process as the dominant mechanism of human/mouse SHM.
\item
We re-examine these results by estimating the false positive rates of their approach.
\item
We find no evidence that templated mutagenesis is a major contributor to SHM.
\end{itemize}

\clearpage 
\section*{Introduction}

Our immune systems generate a highly diverse set of antibodies to protect us from pathogens.
An important part of this process is affinity maturation, which generates high-affinity antibodies for antigens encountered by the immune system.
Affinity maturation is the result of multiple rounds of mutation and selection: mutations are introduced into the rearranged antibody gene by enzymatic processes, and mutations leading to higher-affinity antibodies are selected.

Two major processes are believed to underlie the mutation processes in B cells: classical somatic hypermutation (SHM) and gene conversion (GCV).
Both processes depend on activation-induced cytidine deaminase (AID) \cite{Methot2017-gi}, which creates U:G lesions in the DNA by deaminating deoxycytidine to deoxyuridine.
In SHM, the lesion is resolved by recruiting error-prone repair machinery which can introduce non-templated point mutations at and around the AID-induced lesion.
In GCV, the lesion is repaired using a homologous segment elsewhere in the genome as a donor template, resulting in the homologous tract being copied into the rearranged antibody gene.

In a recent paper, \dea\ \cite{Dale2019-db} propose that new mutation process, called ``templated mutagenesis,'' is also an important contributor to B cell receptor diversification in mice and humans.
In this process, an incompletely-understood mechanism uses the sequence of other germline genes to guide the mutation process at a rearranged germline gene.
One candidate mechanism for this process is gene conversion.
However, templated mutagenesis differs from previous descriptions of gene conversion (in species such as chicken) in that it does not require long stretches of homology between donor and recipient sequences.
Indeed, \dea\ find that templated mutagenesis ``extends into the somatically mutated non-Ig sequences, LAIR1, \emph{gpt}, and $\beta$-globin, despite the lack of overt homology between these genes and the IgHV repertoire.''
For this paper we will simply refer to this newly-hypothesized process as templated mutagenesis.

Although much of the evidence presented by \dea\ was in the form of statistically significant deviations from a simplified null model, the authors suggest that $\sim$50-65\% of mutations in IgH from a collection of human and murine data sets are consistent with templated mutagenesis.
If over half of mutations truly come from templated mutagenesis, B cell repertoire analysis methods will need to be rebuilt.
For example, to estimate the likelihood of a group of mutations that match a template elsewhere in the Ig locus, one must incorporate the respective likelihoods that the group occurred from classical SHM or from templated mutagenesis.
Therefore, any method that relies on estimating mutation probabilities would need to be updated.
This includes all core methods for B cell sequence analysis: germline annotation, lineage tree estimation, selection strength estimation, and validation techniques such as repertoire simulation.
In light of this dependence, accurate rate estimates of templated mutagenesis are crucial.

In this paper, we show that data from human samples \cite{Bornholdt2016-ii} and data from a transgenic mouse model \cite{Yeap2015Sequence} analyzed by \dea\ do not support a high rate of templated mutagenesis.
We do so by re-implementing the software described in \dea, called PolyMotifFinder, which identifies potential templated mutagenesis events by comparing mutated sequences to a pool of potential donor genes.
Using this software, we run a control experiment absent from the original analysis: we calculate the rate of templated mutagenesis using a donor set of simulated genes \emph{not} present in the organism from which the mutated sequences derived.
In this way, we show that the PolyMotifFinder strategy for detecting templated mutagenesis via microhomology has a false positive rate very close to, and in some cases above, the reported positive rate.
This yields an upper bound on the range of the true templated mutagenesis rate; this range is often zero.
We also describe how \dea\ conflate a non-trivial rate of templated mutagenesis with significance estimates for a simplified null model.
In addition, we argue that clustering of mutations is compatible with the classical Neuberger model of SHM and thus evidence of such clustering is not \emph{prima facie} evidence of templated mutagenesis.
We conclude that more evidence is needed if templated mutagenesis should be accepted as an important part of BCR diversification.

\clearpage
\section*{Materials and methods}
Because the original software in \dea\ was not made available as part of the publication, nor was it available upon personal request without a materials transfer agreement restricting its use, we re-implemented the algorithms described in \dea\ as open-source software in Python, a widely available and free programming language.

The PolyMotifFinder algorithm relies on the creation of two matrices.
Given a set of $n$ mutated sequences all deriving from the same germline sequence, the length of which is $p$, and a window size $k$ corresponding to the minimum allowable donor tract length for templated mutagenesis, we create matrices $M$ and $S$, each having $n$ rows and $p$ columns.
$S_{ij} = 1$ if (i) there is a mutation at position $j$ in sequence $i$ and (ii) there exists a window of size $k$ around position $j$ in the $i$th sequence that contains at least two mutations and is represented in the donor set.
That is, $S_{ij}$ is an indicator of position $j$ in sequence $i$ belonging to a pair of mutations consistent with templated mutagenesis.
For $M$, we take $M_{ij} = 1$ if (i) there is a mutation at position $j$ in sequence $i$, (ii) there is at least one other mutation in sequence $i$ whose distance from $j$ is $k-1$ or less, and (iii) that mutation is not part of a pair of mutations that was seen in one of the previous sequences.
In other words, $M$ identifies the set of unique mutations within a length-$k$ window from other mutations.
In \dea, $S$ is referred to as the scoring matrix and $M$ as the mutation matrix.
The templated mutagenesis coverage is then computed as $\sum_{i=1}^n \sum_{j=1}^p M_{ij} S_{ij} / \sum_{i=1}^n \sum_{j=1}^p M_{ij}$.

We note in passing that calculation of the templated mutagenesis coverage depends on the order in which the sequences are processed.
As an example, consider the very abbreviated case where the germline sequence is AAA, the donor set is the single sequence CT (so that $k = 2$), and the mutated sequences are ATT and CTT.
If the sequences are processed as ATT followed by CTT, we will have
\[
  S = \begin{pmatrix} 0 & 0 & 0 \\ 1 & 1 & 0 \end{pmatrix}, \quad M = \begin{pmatrix} 0 & 1 & 1 \\ 1 & 0 & 0\end{pmatrix}
\]
Then $\sum_{i,j} M_{ij} S_{ij}  =1$, $\sum_{i,j} M_{ij}  = 3$, for a templated mutagenesis coverage of $1/3$.

If the sequences are processed in the opposite order, CTT followed by ATT, we have
\[
S = \begin{pmatrix} 1 & 1 & 0 \\ 0 & 0 & 0 \end{pmatrix}, \quad M = \begin{pmatrix} 1 & 1 & 1 \\ 0 & 0 & 0 \end{pmatrix}
\]
Then $\sum_{i,j} M_{ij} S_{ij}  =2$, $\sum_{i,j} M_{ij}  = 3$, for a templated mutagenesis coverage of $2/3$.
Nevertheless, we re-implemented this order-dependent procedure.

PyMotifFinder, our package including a re-implementation of PolyMotifFinder, is available on GitHub (\url{https://github.com/matsengrp/PyMotifFinder}), and our analysis scripts are available at \url{https://github.com/matsengrp/TemplatedMutagenesis-1}.
Our implementation takes as input pairs of naive and mutated sequences along with a donor gene set, the set of potential templates for templated mutagenesis.
It then computes a templated mutagenesis coverage according to the strategy defined above.
Our implementation contains unit tests to verify the accuracy of the algorithm.
The sequence data used in our analyses are available on Zenodo (\url{https://doi.org/10.5281/zenodo.3572361}).
The complete analysis starting from preprocessed data, including generating the figures and tables included in this article, can be reproduced by copy-pasting a handful of commands into a provided Docker container \cite{Boettiger2015-vi} as described in the GitHub repository.
All preprocessing scripts are included with the data on Zenodo.

\subsection*{Sequence data sets}
We analyzed three sets of mutated sequences: one from human subjects, and two from a transgenic mouse model.
The first set of sequences, described in \cite{Bornholdt2016-ii}, corresponds to antibodies to the membrane-anchored Ebola virus glycoprotein trimer.
They were collected from the peripheral B cells of a convalescent donor who survived the 2014 Ebola Zaire outbreak, and will be referred to as the anti-Ebola sequences.
The sequences were downloaded from GenBank using the accession numbers corresponding to the heavy-chain sequences (taken from the supplemental material of \cite{Bornholdt2016-ii}), and both the accession numbers and sequences used are available on Zenodo.

The second and third sets of mutated sequences come from a transgenic mouse model described in \cite{Yeap2015Sequence} which was also investigated by \dea.
Briefly, these sequences come from the B cells of mice that have been genetically engineered with a modified heavy-chain locus.
One chromosome, with the ``productive'' allele, contains the pre-rearranged V region of the 4-hydroxy-3-nitrophenylacetyl (NP)-binding B1-8 antibody (VB1-8).
The other chromosome, with the ``passenger'' allele, contains a sequence consisting of a VB1-8 promoter and leader, followed by a stop codon, followed by either the {\em E. coli} {\em gpt} gene or another copy of VB1-8.
The sequences on both alleles accumulate mutations by somatic hypermutation following the immunization of the mice with NP-chicken gamma globulin.
Since the sequences on the passenger allele cannot be expressed, the SHM patterns on these sequences are unaffected by natural selection, making the system particularly useful for studying SHM.
The analyses presented here use only the passenger {\em gpt} or VB1-8 sequences from this system.
The sequences come from B cells collected from either the Peyer's patches or the spleens of vaccinated mice (six samples from each), and we included sequences taken from both tissue types in our analysis.
These sequences will be referred to as the {\em gpt} sequences and the VB1-8 sequences, respectively.

The {\em gpt} and VB1-8 sequences were downloaded from the Sequence Read Archive (SRP061422).
pRESTO \cite{Vander_Heiden2014-pd} was used to assemble the raw paired-end reads, filter reads to those with an average quality score of at least 20, remove PhiX contamination, and filter to sequences that were seen at least twice.
The script used for this process is available with the data on Zenodo.

\subsection*{Donor gene sets}
For each mutation tract, PolyMotifFinder looks for templated mutagenesis from a provided donor gene set, which contains potential templated mutagenesis donors: if a mutation tract in a mature sequence matches exactly to a region in the donor gene set, the mutation is explainable by templated mutagenesis from that donor set.
We prepared five donor gene sets: a set of human IGHV genes, two sets of mouse IGHV genes, and two ``mock'' sets containing simulated genes homologous to the {\em E. coli} {\em gpt} gene.
Because these simulated {\em gpt} homologs are not present in mouse, we use them as controls as described below.
We refer to these sets as the human IGHV gene set, the mouse IMGT IGHV gene set, the mouse 129S1 IGHV gene set, and the mock {\em gpt} gene sets.

The human IGHV donor set was used to obtain the PyMotifFinder (PyMF) rate estimate of templated mutagenesis in the anti-Ebola sequences.
This set consists of human IGHV gene segments downloaded from IMGT (\url{http://www.imgt.org/vquest/refseqh.html}) in September 2017.
The IMGT label for this set was ``F+ORF+all P,'' corresponding to all functional genes, all open reading frames, and all pseudogene alleles, yielding 466 total segments.
A file containing these segments is available on Zenodo.

The mouse IMGT IGHV gene set was used to obtain the PyMF rate estimate of templated mutagenesis in the VB1-8 and {\em gpt} sequences.
This set consists of mouse IGHV gene segments downloaded from IMGT (\url{http://www.imgt.org/vquest/refseqh.html}) in September 2017.
The IMGT label for this set was ``F+ORF+all P,'' corresponding to all functional genes, all open reading frames, and all pseudogene alleles, yielding 499 total segments.
A file containing these segments is available on Zenodo.

The mouse 129S1 IGHV donor set was used primarily to describe what the mutation spectrum would look like under a templated mutagenesis model in the transgenic mouse system described in \cite{Yeap2015Sequence}.
While these mice belonged to strain 129P2, the 129P2 heavy-chain locus has not been sequenced; instead, we used the set of IGHV genes present in the closely related strain 129S1, published in \cite{Retter2007-zr}.
We note that the gene set is incomplete, with only the 3' half of the IGHV locus sequenced.
Since this gene set was used primarily to describe the likelihood of mutations in a model of templated mutagenesis and not to get at the rate of templated mutagenesis, we decided that a partial set of genes that match closely those found in in the actual system was an appropriate choice.
In particular, we believe it is better than the alternative of using the mouse IGHV genes taken from IMGT, which include several times as many genes as are present the 129P2 genome.
The sequenced and annotated region of the 129S1 genome was downloaded from GenBank (\url{https://www.ncbi.nlm.nih.gov/nuccore/126349412}).
The V gene segments were extracted using a custom Python script available on Zenodo.

The two mock {\em gpt} donor sets were used to estimate the false positive rate of the PolyMotifFinder strategy with the human IMGT donor set and the false positive rate of the PolyMotifFinder strategy with the mouse IMGT donor set.
For a mock donor set to give a good estimate of the false positive rate of the PolyMF strategy, the mock set should have approximately the same homology structure as the donor set used by PolyMF.
We created one such set for the human IMGT IGHV donor set and one for the mouse IMGT IGHV donor set.
In each case, we aligned the sequences in the gene set using MUSCLE version 3.8.31 \cite{Edgar2004-rk} and inferred a phylogenetic tree on the sequences using FastTree version 2.1.7 \cite{Price2010-cy}.
We then used pyvolve \cite{Spielman2015-gl} to simulate a new set of sequences from the estimated phylogenetic tree.
In each case, the {\em gpt} sequence was the root, and the sequences were simulated according to a continuous-time Markov process along the estimated phylogeny.
We used the GY94 codon model \cite{Goldman1994-to} with parameters $\alpha = .98$, $\beta = .65$.

All of the donor sets are available on Zenodo, along with the scripts used to extract the IGHV gene segments from the 129S1 genome, the scripts to align and create trees from the human IGHV genes and {\em gpt} genes, and the script to create the two mock {\em gpt} donor sets.

\subsection*{Germline annotation and mutation calling}
To use PyMF on the anti-Ebola, VB1-8, and {\em gpt} sequences, we needed to identify the mutations and the naive sequences.
For the anti-Ebola and VB1-8 sequences, germline sequences and mutations were identified using partis version 0.13.0 \cite{Ralph2016-hj} with default germline V, D, and J gene sets.
These sets comprise curated subsets of the germline genes in IMGT: excluded are genes that are biologically implausible (e.g. on the wrong chromosome, non-functional, lacking the conserved cysteine) or otherwise considered inaccurate \cite{Wang2008-ic}.

For the {\em gpt} sequences, partis was run using a modified set of germline genes.
The reference sequence for the passenger {\em gpt} gene (obtained via personal communication with Dr.\ Leng-Siew Yeap) was
\begin{verbatim}
CTTTCTCTCCACAGGTGTCCACTCCCAGGTCCAACTGTAGTAGATGAGCGAAAAATACATCGTCACCTGGGACAT
GTTGCAGATCCATGCACGTAAACTCGCAAGCCGACTGATGCCTTCTGAACAATGGAAAGGCATTATTGCCGTAAG
CCGTGGCGGTCTGGTACCGGGTGCGTTACTGGCGCGTGAACTGGGTATTCGTCATGTCGATACCGTTTGTATTTC
CAGCTACGATCACGACAACCAGCGCGAGCTTAAAGTGCTGAAACGCGCAGAAGGCGATGGCGAAGGCTTCATCGT
TATTGATGACCTGGTGGATACCGGTGGTACTGCGGTTGCGATTCGTGAAATCTGCAGTGACGCGCCCACTCTCAC
AGTCTCCTCAGGTGAGTCCTTACAACCTCTCTCTT
\end{verbatim}
In the reference sequence, positions 44 through 351 correspond to the first 308 nucleotides of the {\em gpt} gene and the remainder are linkers.
To align and call mutations from the germline sequence, partis requires a set of germline V, D, and J gene segments.
For the V gene segment, we used the first 308 nucleotides of the {\em gpt} gene, i.e., the portion of the {\em gpt} gene inserted into the mouse germline.
For the D and J gene segments, we used arbitrary ``fake'' gene segments: {\tt AAAAAAAAAA} for the D gene segment and {\tt GGGGGGGGGG} for the J gene segment.
We appended these segments to the end of each sequence so that each input sequence to partis ended with {\tt AAAAAAAAAAGGGGGGGGGG}.
When used this way, partis aligns the {\em gpt} portion of the sequence to the portion of the {\em gpt} gene included in the reference, aligns the added suffix to the fake D and J gene segments, and treats the linker region following the end of the {\em gpt} gene as a VD insertion.
We validated the results by checking that each inferred ``V'' sequence length was correct, that the inferred mutation rate in the ``V'' gene region was not too high, that the inferred VD insertion lengths were correct, and that the VD insertion sequences corresponded to the linker portion of the naive {\em gpt} sequence (positions 352 through 410 in the sequence above).
Therefore, all mutations identified by partis are located on the {\em gpt} portion of the sequence with none in the linker sequence.
This is the desired behavior for our analysis because we are looking for regions of microhomology in the {\em gpt} sequence and we do not wish to analyze mutations that occurred in the linkers.

\subsection*{Model for mutation probabilities due to templated mutagenesis}
To investigate whether templated mutagenesis could explain the observed mutations, we constructed a simple statistical model for templated mutagenesis.
The model assumes a uniform probability distribution over the possible templated mutagenesis donors, so that each donor is equally likely to have provided the template for a given templated mutation event.
For each mutated site we identified the three possible mutations: the mutation that actually occurred and the two mutations that did not occur.
For each of the three mutations, we identified all ways of aligning a donor gene to a mutation-containing region so that the two match in at least $k$ bases around the mutation.
We defined this set as the set of potential templated mutagenesis donors, and we modeled mutation due to templated mutagenesis as a uniform draw from this set of donors.
In this model, the probability of seeing the observed mutation from a templated mutagenesis event is the number of donors containing the observed mutation divided by the total number of donors.
That is,
\begin{equation}
P(\text{observed mutation} \mid \text{templated mutation event}) = \frac{n_{\text{obs}}}{n_{\text{obs}} + n_{\text{unobs}}}\label{Eq:prob-given-gcv}
\end{equation}
where $n_{\text{obs}}$ is the number of donors matching the observed mutation and $n_{\text{unobs}}$ is the number of donors matching one of the two possible unobserved mutations.
We can compute these probabilities for any donor gene set to evaluate how well it explains the observed pattern of mutations.

\subsection*{Upper bound on the rate of templated mutagenesis}

Given a bound on the false positive rate of PyMF and an estimate by PyMF of the rate of templated mutagenesis, we can compute an upper bound on the true rate of templated mutagenesis.
To do this, we use a simple mixture model in which we assume that mutations arise either by templated mutagenesis or by classical SHM.
We first define the following quantities, all of which take values in $[0,1]$:
\begin{enumerate}
\item The false positive rate, $\text{FPR}$, is the probability that PyMF classifies a mutation due to classical SHM as explainable by templated mutagenesis.
\item The true positive rate, $\text{TPR}$, is the probability that PyMF classifies a mutation due to templated mutagenesis as explainable by templated mutagenesis.
\item The positive rate, $\text{PR}$, is the overall probability that PyMF classifies a mutation as explainable by templated mutagenesis.
\item $p_{\text{shm}}$ is the true proportion of classical SHM events. Since a mutation can be due to either SHM or templated mutagenesis but not both, $1-p_{\text{shm}}$ is the true proportion of templated mutagenesis events.
\end{enumerate}
Then the overall probability that PyMF classifies a mutation as explainable by templated mutagenesis is
\[
\text{PR} = p_{\text{shm}} \times \text{FPR} + (1 - p_{\text{shm}}) \times \text{TPR}.
\]
If we assume that $\text{TPR} = 1$ so that all true templated mutagenesis events are correctly classified by PyMF as due to templated mutagenesis (i.e., PyMF has sensitivity 1) and that $\text{FPR} \ge b$ for some lower bound $b$, we can rearrange the expression above to conclude that $p_{\text{shm}} \ge \frac{1 - \text{PR}}{1 - b}$.
Equivalently, the rate of templated mutagenesis would be at most $1 - \frac{1-\text{PR}}{1 - b}$.
If we do not specify a true positive rate, the upper bound for the rate of templated mutagenesis becomes
\begin{equation}
1 - \frac{\text{TPR} - \text{PR}}{\text{TPR} - \text{FPR}}\label{Eq:upperbound},
\end{equation}
assuming that $\text{TPR} > \text{FPR}$.
If $\text{TPR} < \text{FPR}$, we cannot obtain an upper bound.
We apply \eqref{Eq:upperbound} to obtain upper bounds on the rate of templated mutagenesis in mice and humans.

\subsection*{Hypothesis testing and confidence intervals}

The {\em gpt} sequences have a grouped structure: each mutated sequence comes from one of 12 tissue samples from 6 different organisms, and so it is inappropriate to model them as independent and identically distributed.
Hypothesis testing and confidence interval construction for the {\em gpt} sequences was therefore performed using a mixed effects model fit with the {\tt lme4} package \cite{Bates-lme4} in R \cite{Rcore-2017}.
For each value of $k$ (the minimum donor tract length) and each reference sequence, we modeled the probability of a mutation given templated mutagenesis using a mixed model with a random effect for tissue sample.
Confidence intervals were plotted as the fitted value in the mixed model plus or minus two standard errors.
To test whether templating from the IGHV genes could explain the observed mutations better than templating from the {\em gpt} genes, we computed the probability of each mutation under the model of templated mutagenesis by IGHV genes and templated mutagenesis by {\em gpt} genes.
We then computed the difference between the probability of the mutation under the IGHV templating model and the {\em gpt} templating model.
Under the null hypothesis that the two models are equally good at explaining the observed mutations, these differences should have mean zero.
As before, since the mutations have a grouped structure, we tested this null hypothesis using a mixed model with a random effect for tissue sample.

\clearpage
\section*{Results}

\subsection*{PyMotifFinder identifies a high fraction of mutations as explainable by templated mutagenesis}

To verify that our re-implementation of PolyMotifFinder was comparable to the version presented by \dea, we ran PyMF on the {\em gpt} and VB1-8 sequences described in \cite{Yeap2015Sequence} with the mouse IGHV gene set, as well as the anti-Ebola sequences described in \cite{Bornholdt2016-ii} with the human IGHV gene donor set.
We found slightly higher but comparable rates of mutations explainable by templated mutagenesis in the {\em gpt} sequences: of the mutations within 8 nucleotides of each other, 60-75\% had 8-mer templates in the mouse V genes (Figure \ref{Fig:fpr}), consistent with the initial report.
We found a higher rate of mutations explainable by templated mutagenesis in the anti-Ebola and VB1-8 sequences: of the mutations within 8 nucleotides of each other, 73\% of the anti-Ebola sequences had 8-mer templates in the human V genes and 75\% of the VB1-8 sequences had 8-mer templates in the mouse V genes.
The discrepancy is likely due to differences in gene filtering, germline annotation, and mutation calling between our respective pipelines.
Indeed, when we perform stricter filtering to our donor gene sets by removing open reading frame and pseudogene sequences, we obtain rates estimates of 42\% for humans and 62\% for mice, which are very close to the \dea\ estimates.
We discuss this in detail in Section "Consistent results using filtered donor gene sets".

\subsection*{{\em gpt} sequences can be used to estimate the PolyMotifFinder false positive rate}
To investigate the false positive rate of the PolyMotifFinder strategy, we ran PyMF on the set of somatically mutated {\em gpt} sequences described in the "Sequence data sets" section using a mock donor set of simulated {\em gpt} homologs that are not present in B cells.
Since mutations could not have arisen from copying over templates from our mock set, any inference of templated mutagenesis events identified by the method must be a false positive.
For the rate of false positives provided by the mock donor gene set to provide a good estimate of the true false positive rate, the mock donor gene set should be constructed so that the probability that an SHM-induced mutation in a {\em gpt} sequence matches a member of the mock donor gene set is close to the probability that an SHM-induced mutation in a real antibody sequence matches one of the IGHV genes.
For this to hold, the distribution of molecular divergences among the genes in the mock donor set should match the distribution of divergences among the real donor gene set.

Our mock donor gene sets were created to have these properties.
To verify this, we estimated phylogenies for the two mock donor gene sets and the two IMGT IGHV gene sets and computed the divergences between the roots and the leaves in each.
The divergence distribution in the mock {\em gpt} set based on the mouse IMGT gene set resembled the divergence distribution in the mouse IMGT gene set (mean divergence $.48$ and $.48$, respectively).
The same holds with the mock {\em gpt} set based on the human IMGT gene set and the human IMGT gene set (mean divergence $.41$ and $.4$, respectively).
For a more complete description of the divergences, five-number summaries of the divergences in each of the four gene sets are given in Supplemental Table I.

Finally, we note that in the {\em gpt} system, we expect all of the mutations to be introduced by SHM.
Although \dea\ analyze these sequences and suggest that templated mutagenesis could be occurring, the lack of homology between the {\em gpt} gene and the V genes makes it {\em a priori} unlikely that these mutations are introduced by templated mutagenesis.
We address the potential contribution from small micro-homologies (matches of fewer than 10 bases) between the {\em gpt} gene and the IGHV genes due to chance sequence similarity in Section ``Evidence that mutations in {\em gpt} sequences are not due to templating from V genes''.

\subsection*{{\em gpt} analysis demonstrates that the MotifFinder methodology has a high false positive rate}

\begin{figure}
\begin{center}
\includegraphics[width=.85\textwidth]{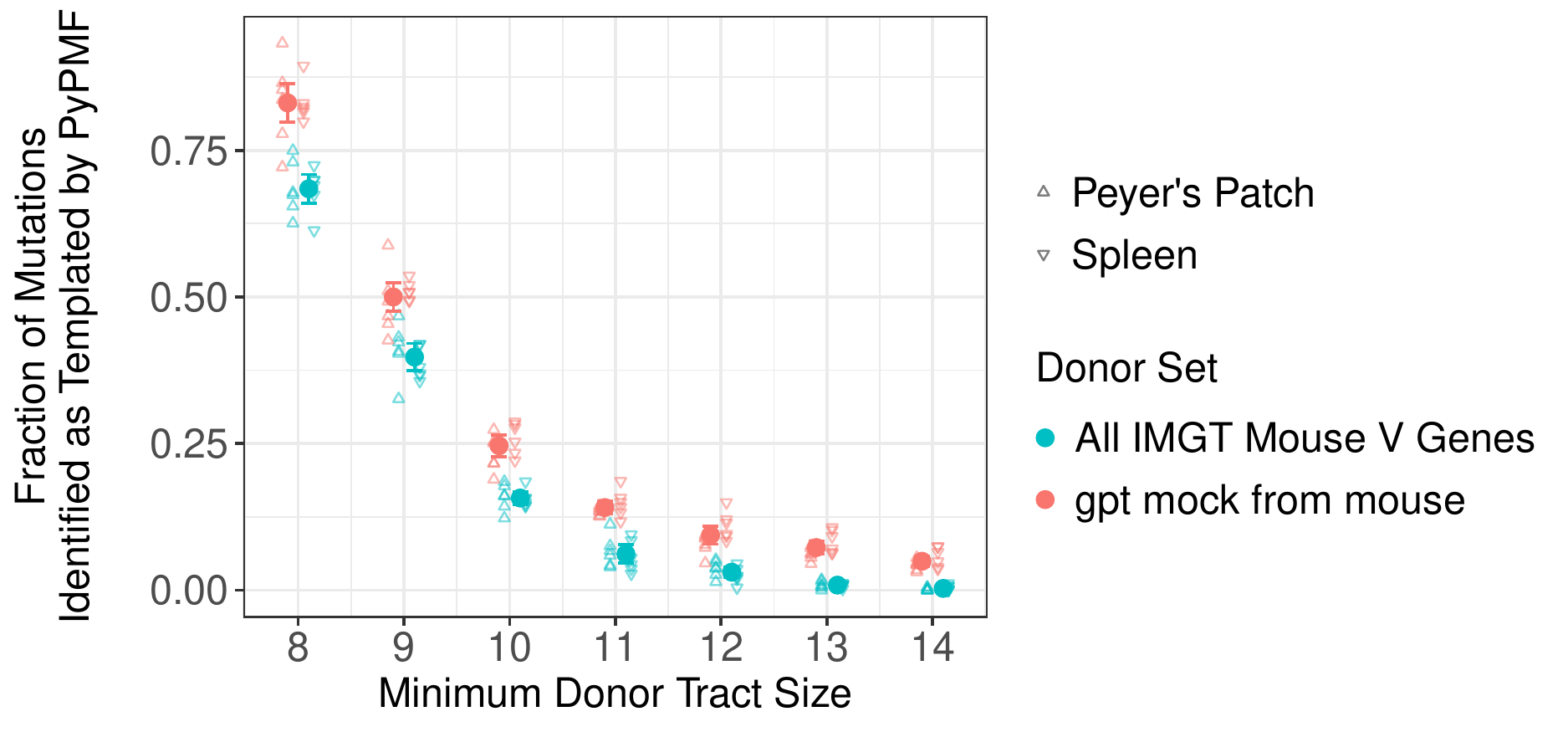}
\end{center}
\caption{%
  Red points represent the fraction of mutations in the {\em gpt} gene explainable by templates in the mock donor set of simulated {\em gpt} homologs, which are not present in the mouse germline and are not available as templates for templated mutagenesis (i.e., the FPR).
  Blue points represent the fraction of mutations in the {\em gpt} gene explainable by templates in the mouse IMGT IGHV gene donor set, which are potentially present in the mouse germline and available as templates.
  For each tract length and donor set, the filled circle and error bar represents the overall estimate of the probability of a mutation being explainable by templated mutagenesis plus or minus two standard errors.
  We see that the FPR of PyMF is larger on average than the PyMF estimate of the fraction of mutations explainable by templated mutagenesis.
  Points corresponding to samples from Peyer's patches and spleen are offset slightly to the left and right, respectively, to facilitate comparison and to avoid overplotting.
  This analysis was performed once on data from six individual mice, with two replicates per mouse corresponding to samples from Peyer's patches and spleen, yielding 12 total samples.%
}
\label{Fig:fpr}
\end{figure}

To estimate the false positive rate of PyMF with the human IMGT IGHV gene set and the false positive rate of PyMF with the mouse IMGT IGHV gene set, we ran the algorithm on the {\em gpt} sequences with the corresponding mock {\em gpt} donor gene set.
Since the donor set of simulated {\em gpt} homologs is not present in the mouse, they cannot have been used as templated mutagenesis donors, and any mutation PyMF identifies as explainable by templated mutagenesis from this set is a false positive.
We ran PyMF with minimum donor tract length ranging from 8 to 14, and we found false positive rates that were on the same order as the PolyMF rates obtained when mutated sequences were run against real donor sets.
The absolute false positive rates were particularly high for minimum donor tracts of 8 and 9 (Figure \ref{Fig:fpr}).
The average false positive rate was 83\% for a donor tract of 8, 50\% for donor tracts of size 9, and 25\% for donor tracts of size 10.
This rate falls dramatically as the minimum donor tract size increases, dropping to 5\% for donor tracts of size 14.
This suggests that the PolyMotifFinder strategy has a large false positive rate for small values of $k$, classifying more than 50\% of mutations as explainable by templated mutagenesis from genes that were not present in the mouse when $k = 8$ or $k = 9$.

\subsection*{Evidence that mutations in {\em gpt} sequences are not due to templating from V genes}

One might explain the high false positive rate of PolyMotifFinder by saying that in spite of the overall lack of homology between {\em gpt} genes and V genes, the mutations in the {\em gpt} sequence were actually introduced by templating from very small homologous tracts in the mouse V genes.
To check this possibility, we ran PyMF on the {\em gpt} sequences with the mouse IMGT IGHV genes as a donor set.
We found that many of the mutations could be explained by templating from very small homologous tracts in the mouse V genes, corresponding to the findings of \dea.
For example, nearly 60\% of the mutations had a template of size 8 in the V gene set and about 40\% of the mutations had a template of size 9.
However, the percentage approaches zero as template size increases, and for every template size the average proportion of mutations explainable by templating from {\em gpt} sequences is higher than the average proportion of mutations explainable by templating from V genes (Figure \ref{Fig:fpr}).
In fact, the proportion of mutations explainable by templating from the V genes is very close to zero for templating by tracts of size 11 or greater, while the proportion explainable by templating from the mock {\em gpt} donor set remains non-neglibible.

\begin{figure}
\begin{center}
\includegraphics[width=.85\textwidth]{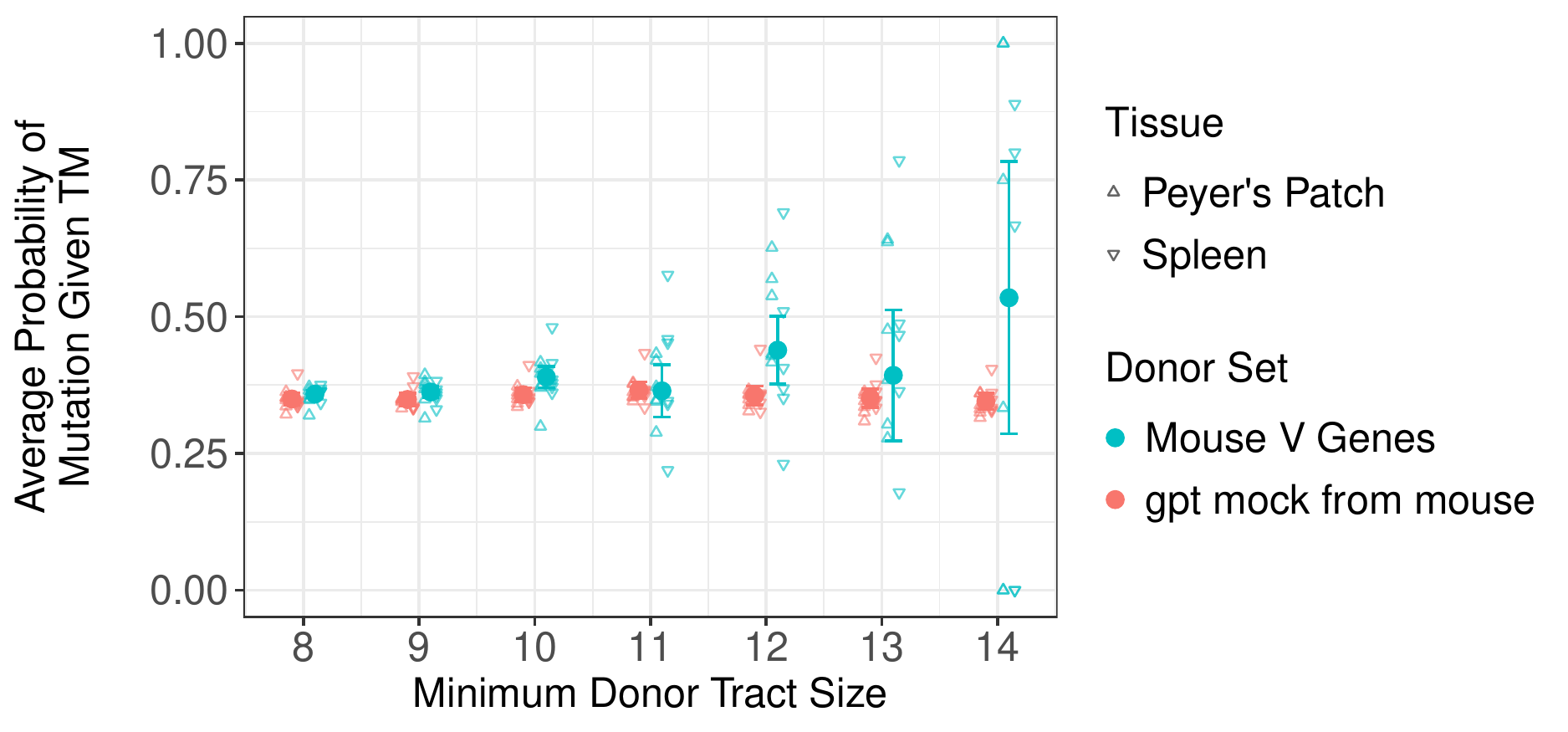}
\end{center}
\caption{%
Average probability of the observed mutations under a templated mutagenesis model, either templating from {\em gpt} genes or templating from the set of 129S1 V genes.
Each point corresponds to one sample taken from either spleen or Peyer's patches, so that the average is computed over all sequences in a given sample.
This analysis was performed once on data from six individual mice, with two replicates per mouse corresponding to samples from Peyer's patches and spleen, yielding 12 total samples.%
}
\label{Fig:prob-given-gcv}
\end{figure}

We next formally evaluated the plausibility that the mutations in the {\em gpt} sequences were introduced by templated mutagenesis from the IGHV genes present in the mouse.
To do so, we computed the probabilities of the {\em gpt} mutations via templated mutagenesis from 129S1 IGHV donor gene set and the probabilities of the {\em gpt} mutations via templated mutagenesis from the mock donor set of simulated {\em gpt} homologs, using the uniform-across-donors probability model specified by Equation \eqref{Eq:prob-given-gcv} for both cases.
These models encode our intuition that if the mutations really were templated from a donor gene set, the observed mutation spectrum should be biased towards bases that are represented more frequently in potential donors from that gene set.
As an example, suppose that we are considering one mutation in the {\em gpt} sequence from A to T.
If all of the potential templated mutagenesis donors in the V gene set would lead to a mutation from A to T and all of the templated mutagenesis donors in the {\em gpt} gene set would lead to a mutation from A to C, the observed A to T mutation is explained better by templating from the V genes than by templating from the {\em gpt} genes.
We can fit one model using donors from the mouse V genes and another model using donors from the {\em gpt} genes and compare how well each model explains the data: if templated mutagenesis from mouse V genes were really occurring, we would expect the V gene model to fit the data better than the {\em gpt} model.
If this is not true, it suggests that both the V gene and the {\em gpt} inferences are spurious, as the {\em gpt} donor genes are not actually present in the mouse.

For each sample, we computed the average probability of the mutations in the {\em gpt} sequences given templated mutagenesis from the mouse IGHV gene donor set and the {\em gpt} donor set for tract sizes ranging from 8 to 14.
We found that these numbers were comparable for the {\em gpt} donor set and the mouse IGHV gene donor set, as shown in Figure \ref{Fig:prob-given-gcv}.
As described in the Methods section, we used a mixed effects model to test for a difference in the expected probabilities of mutation due to templating from the {\em gpt} donor set and the mouse IGHV gene donor set.
The resulting $p$-values were: $p = .059, .054, .115, .202, .001, .213,$ and $.249$ for $k = 8$ through $14$, respectively.
This indicates that the mutations in the {\em gpt} sequences do not tend to look any more like the mouse IGHV gene donor set than they do like the {\em gpt} donor set.
Because the {\em gpt} donor set was not present in the mouse, we believe that it is unlikely that the mutations in the {\em gpt} sequences were introduced by templating from the mouse IGHV genes.

\begin{figure}
\begin{center}
\includegraphics[width=.85\textwidth,page=1]{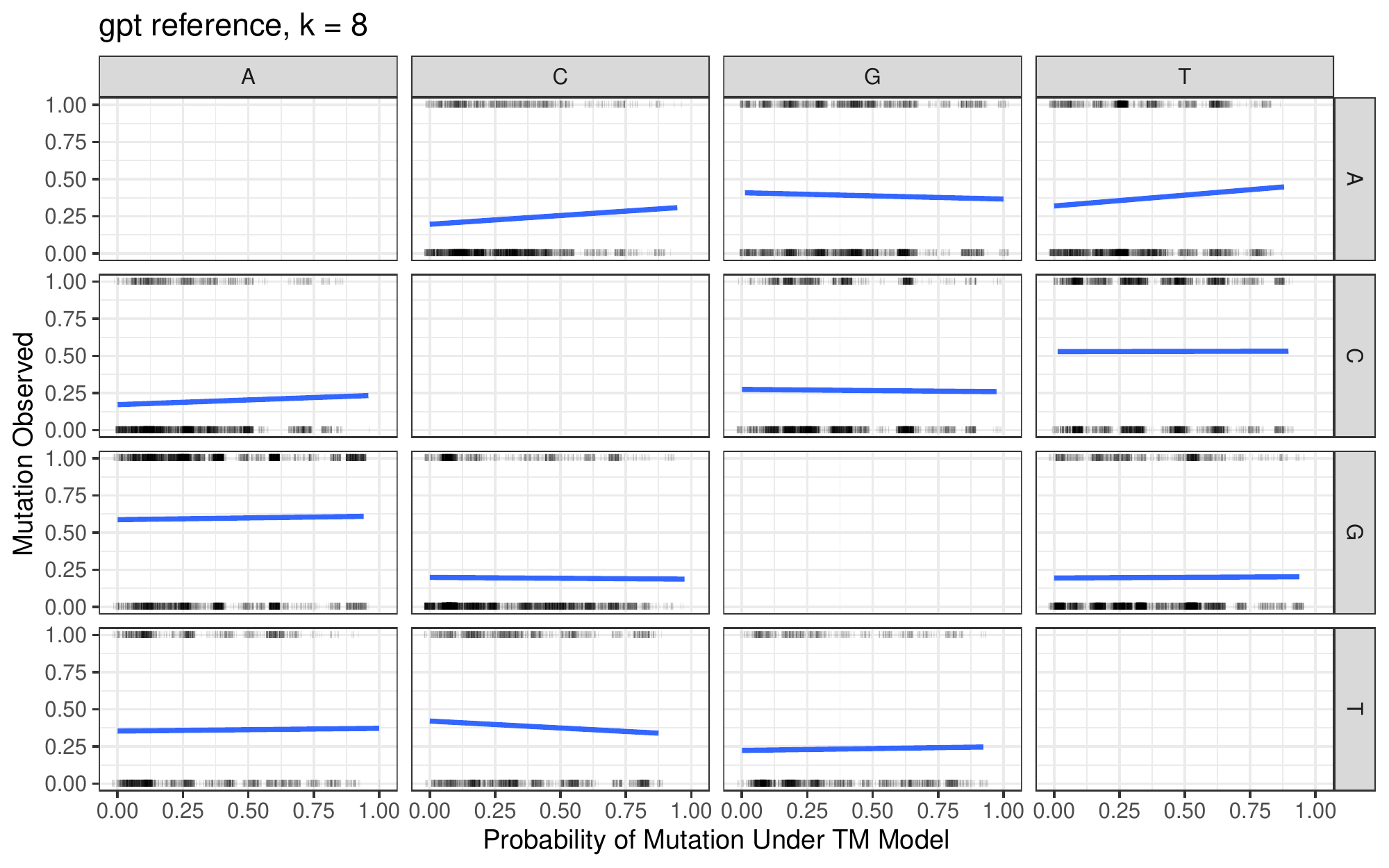}
\includegraphics[width=.85\textwidth,page=2]{per_base_obs_vs_exp}
\end{center}
\caption{\small Each subplot displays whether a mutation was observed (on the $y$-axis) versus its probability under the templated mutagenesis model (on the $x$-axis).
A $y$-value of one means the mutation was observed, and a $y$-value of zero means the mutation was not observed. For each mutation, the germline base is indicated by the row name, and the target base indicated by the column name.
The lines are linear smoothers.
We do not observe any consistent and significant trend to these lines, indicating that templated mutagenesis has not contributed to the observed sequence changes in the {\em gpt} sequence data set.
This analysis was performed once on data from six individual mice, with two replicates per mouse corresponding to samples from Peyer's patches and spleen, yielding 12 total samples.
}
\label{Fig:obs-vs-exp-per-base}
\end{figure}

To further investigate whether the mutations could have arisen due to templated mutagenesis from the mouse V genes, we asked whether mutations that had a higher probability under the templated mutagenesis model were observed more frequently.
For each mutation from germline base $b_1$, we computed the probability of mutation from $b_1$ to any of the other three bases at that position under the templated mutagenesis model.
We then asked whether target bases that had a higher probability under the templated mutagenesis model were observed more frequently.

We found that mutations with higher probabilities under the templated mutagenesis model were not observed any more frequently than mutations with low probabilities under the model (Figure~\ref{Fig:obs-vs-exp-per-base}).
This finding held true both for the model of templated mutagenesis from {\em gpt} genes and from the 129S1 IGHV genes.
To formally test whether mutations with high probabilities occurred more frequently, we performed independent logistic regressions for each pair of germline and target base.
The response variable was an indicator of whether the observed mutation was the target base, and the predictor variable was the probability of mutation to the target base under the templated mutagenesis model.
In each case, we found that the slope in the model was non-significant at the .05 level, indicating that the templated mutagenesis model did not help to explain the observed pattern of mutations.
This analysis provides further evidence that the mutations in the {\em gpt} sequences did not arise by templated mutagenesis from the IGHV genes present in the mouse.

\subsection*{Upper bounds on the rate of templated mutagenesis}
We combined PyMF's estimate of the rate of templated mutagenesis with our estimate of PyMF's false positive rate to obtain an approximate upper bound on the true rate of templated mutagenesis in mice (using the VB1-8 sequences) and in humans (using the anti-Ebola sequences).
Plugging in PyMF's estimates of the rate of templated mutagenesis in the VB1-8 sequences and our estimates of PyMF's false positive rates from the {\em gpt} sequences to Equation~\ref{Eq:upperbound}, we found upper bounds on the rate of templated mutagenesis in this system ranging from 0 (in cases where our estimate of the false positive rate exceeds the rate at which PyMF identified templated mutations) to $.1$, depending on the value of $k$ and the assumed true positive rate (Table \ref{Tab:bound}, top panel).
The largest upper bounds were obtained at $k = 8$.
For the human anti-Ebola sequences, we found upper bounds on the rate of templated mutagenesis ranging from $0$ to $.12$, with the numbers again varying based on the value of $k$ and the assumed true positive rate.
In this case, the largest upper bounds is obtained at the largest value of $k$, $k = 14$, and in general the larger values of $k$ correspond to larger upper bounds.
However, note that, since these estimates are upper bounds of the true rates in both humans and mice, they are consistent with a rate of zero.

\begin{table}[ph!] \centering
 Mice:\\
 \begin{tabular}{l | l l l l l l}
 $k$&PyMF rate&PyMF FPR&UB (1)&UB (.99)&UB (.95)&UB (.9)\\\hline
 8&0.79&0.83&0&0&0&0\\
 9&0.54&0.5&0.08&0.08&0.09&0.1\\
 10&0.28&0.25&0.05&0.05&0.05&0.05\\
 11&0.15&0.14&0.01&0.01&0.01&0.02\\
 12&0.11&0.09&0.01&0.01&0.01&0.02\\
 13&0.07&0.07&0&0&0&0\\
 14&0.06&0.05&0.01&0.01&0.01&0.01
 \end{tabular}

 \vspace{2em}

 Humans:\\
 \begin{tabular}{l | l l l l l l}
 $k$&PyMF rate&PyMF FPR&UB (1)&UB (.99)&UB (.95)&UB (.9)\\\hline
 8&0.73&0.78&0&0&0&0\\
 9&0.44&0.43&0.02&0.02&0.02&0.02\\
 10&0.26&0.2&0.07&0.08&0.08&0.09\\
 11&0.18&0.11&0.09&0.09&0.09&0.1\\
 12&0.15&0.07&0.09&0.1&0.1&0.11\\
 13&0.15&0.05&0.1&0.1&0.11&0.11\\
 14&0.13&0.03&0.1&0.11&0.11&0.12
 \end{tabular}
 \caption{Upper bounds (UB) on the rate of templated mutagenesis in the VB1-8 (top) and the anti-Ebola sequences (bottom) computed for a range of tract lengths $k$ and sensitivities.
  $k$ denotes tract length, PyMF rate is the naive PyMF estimate of the rate of templated mutagenesis, PyMF FPR is the PyMF false positive rate, UB denotes upper bound, and the number in parentheses denotes the assumed sensitivity (true positive rate) of PyMF.}
 \label{Tab:bound}
 \end{table}

We caution against taking these numbers as definitive as we do not know the true positive rate of PyMF, and they require that our estimate of the false positive rate of PyMF is a lower bound.
However, they attempt to correct the observed rates using false positive rate estimates, and in particular show that templated mutagenesis does not occur at a high rate unless PyMF misses many true templated mutagenesis events.

\subsection*{Consistent results using the reverse complementary strand}

We also tested whether templated mutagenesis could be occurring from the reverse complementary strand.
To this end, we repeated all the analyses with the reverse complements added to the donor gene sets.
The results are shown in Supplemental Figures 1 and 2 and Supplemental Table II, and were qualitatively similar to those with the original gene sets.
The rate estimates were slightly higher because of the larger size of the donor sets (Supplemental Figure 1).
The average probability of the observed mutations given templated mutagenesis from the {\em gpt} genes and their reverse complements remained about the same as the average probability of the observed mutations given templated mutagenesis from the IGHV genes and their reverse complements (Supplemental Figure 2).
The upper bounds on the rate of templated mutagenesis also remained low when the reverse complements were included in the donor gene sets (Supplemental Table II).

\subsection*{Consistent results using filtered donor gene sets}

We obtained positive rate estimates of 73\% for humans and 79\% for mice (Table \ref{Tab:bound}) when applying the PolyMotifFinder strategy with $k = 8$ using our chosen donor gene sets and BCR sequence datasets as discussed in the Materials and Methods section.
In their analysis, \dea\ estimate this positive rate to range to be approximately $50 - 65$\% when applying the same strategy to their chosen donor gene sets and BCR sequence datasets.
While they describe the five different BCR sequence datasets used to obtain these estimates, two things remain unclear.
First, it is not obvious how they extracted the $50-65$\% range from the data displayed in their Figure 5(I), which seems to show positive rate estimates roughly ranging from $\approx$ 30\% to $\approx$ 90\% and whose rate estimates seem to depend on the dataset in question.
Secondly, they do not describe the exact donor gene sets obtained from IMGT.
The construction of the donor sets is crucial since the number of genes in the donor set influence the positive rate estimates: adding more templates to the donor set can only increase the number of PolyMotifFinder hits since there will be more chances to observe a match.

To address these discrepancies, we re-ran both of the VB1-8 and anti-Ebola analyses using a more restricted donor gene set in each case.
Specifically, we filtered out all open reading frame (ORF) and pseudogene (P) sequence reads from the respective IMGT sets, which led to a 31.7\% decrease in potential donors for the VB1-8 sequences and a 44.9\% decrease in potential donors for the anti-Ebola sequences.
We obtained positive rate estimates of 42\% for humans and 62\% for mice for $k = 8$.
Detailed tables of rate estimates using the filtered donor sets, analogous to Table \ref{Tab:bound}, can be found in the main Github respository (\url{https://git.io/Jfl6t}). 
Between the collective full and restricted analyses for mice and humans, our positive rate estimates range from $42-79$\%, which contains the $50-65$\% range proposed by \dea.
More importantly, our estimates on the upper bound of templated mutagenesis events remain highly similar between the full and restricted analyses, demonstrating the robustness of our methodology to the particular choice of donor gene set.

\subsection*{A small $p$-value for a simplified null model does not imply a non-trivial effect size for the rate of templated mutagenesis}

Finally, we point out that our estimates of templated mutagenesis occurring at a low rate are in fact compatible with the large values of Stouffer's $Z$ and the correspondingly small $p$-values obtained in \dea.
These authors compare the rate of templated mutagenesis to the rate obtained using a simplified null model (called RandomCheck) in which, conditional on the locations of the mutations, the mutation identity at each location is independent of the other locations and follows a fixed distribution taken from previous studies.
This model is a simplification of the classical Neuberger model of somatic hypermutation in many ways.
In the Neuberger model, lesions introduced by AID can be resolved by one of three pathways, each of which leads to a repair by a different set of enzymes.
The likelihood of each pathway being recruited to repair the lesion depends on nucleotide context, and each pathway is assumed to have its own unique, context-dependent mutation profile \cite{Rogozin2001-yp}.
The result is that the mutations are not independent and identically distributed conditional on the germline base, in contrast with the assumption of RandomCheck, which was used to compute $p$-values and Stouffer's $Z$ in \dea.
Aside from issues of independence, the overall mutation profile taken from the literature is exceedingly unlikely to be {\em exactly} correct, and, given enough samples, any consistent hypothesis testing framework will confidently identify even small differences between the true mutation profile and the one drawn from the literature.

\begin{figure}
\includegraphics{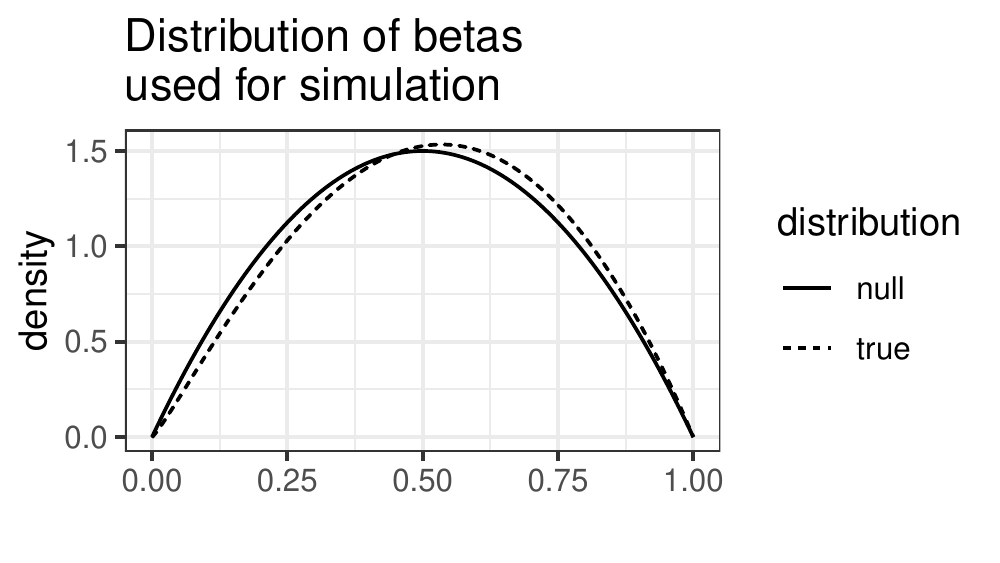}\\
\includegraphics{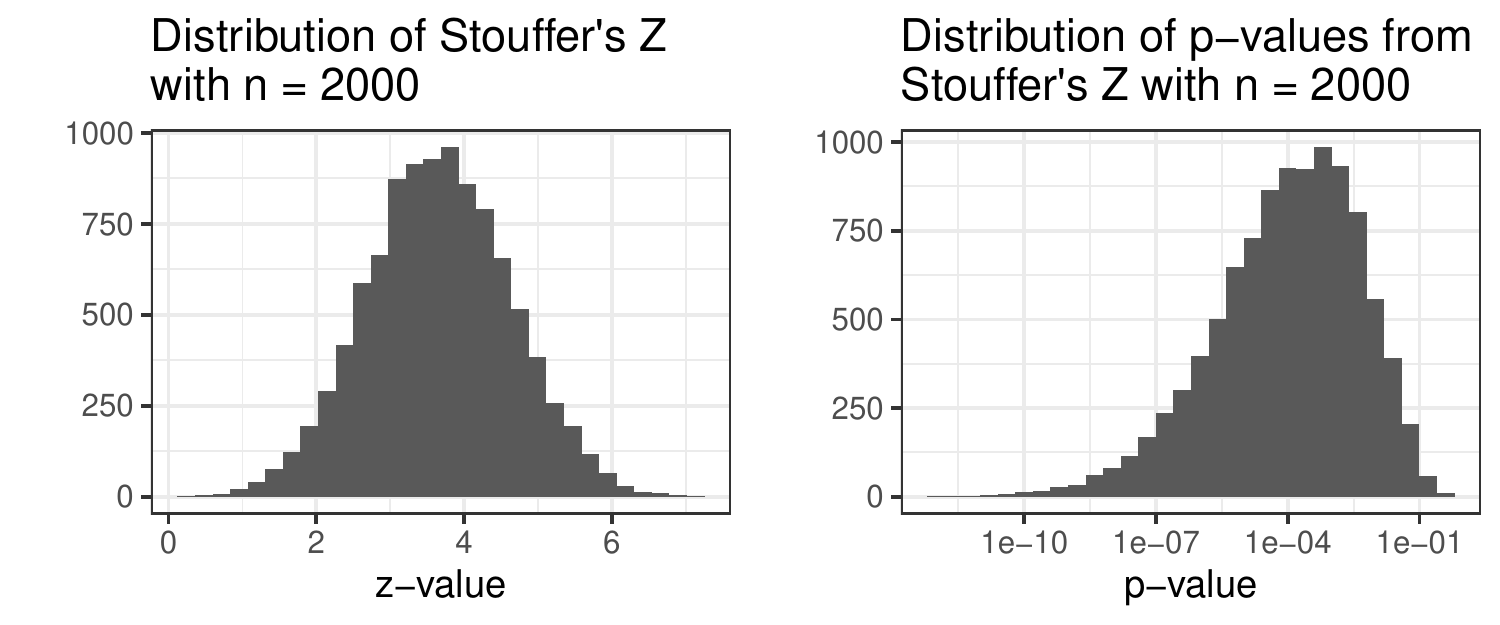}
\caption{%
  Top: Densities of the distributions used in the simulations of Stouffer's $Z$.
  Samples coming from the ``true'' distribution (dashed line) are tested against the hypothesis that they come from the null distribution (solid line).
  Bottom: Distributions of Stouffer's $Z$ statistics (left) and $p$-values (right) for the true and null distributions in the top panel for 10,000 simulation trials. In each trial, the Stouffer's $Z$ value is aggregated over 2,000 independent tests, which is about the same as the number of trials aggregated by \dea.
}
\label{Fig:two-betas}
\end{figure}

To demonstrate that even a slightly misspecified model can lead to extreme values of Stouffer's $Z$ and highly significant $p$-values, we performed a small simulation study.
We suppose that the fraction of mutations explainable by templated mutagenesis in the ``true'' model is drawn from a beta distribution with mean .518 and variance .048, shown as a dashed line in Figure \ref{Fig:two-betas}.
In the null model, the fraction of mutations explainable by templated mutagenesis is drawn from a beta distribution with mean .5 and variance .05, shown as a solid line in Figure \ref{Fig:two-betas}.
We simulate 2,000 values (corresponding to 2,000 \emph{gpt} sequences analyzed) for the fraction of mutations explainable by templated mutagenesis, construct $Z$-values from the hypothesis test that these values come from the null distribution, and finally compute Stouffer's $Z$ from the collection of 2,000 $Z$ values.
We performed this procedure 10,000 times, yielding a distribution of 10,000 Stouffer's $Z$ values.

In this simulation, the values of Stouffer's $Z$ were centered around 3.65 with a standard deviation of .97.
The corresponding $p$-values had a median value of $1.3 \times 10^{-4}$.
10\% of the $p$-values were smaller than $4.2 \times 10^{-7}$, and 90\% were smaller than $7.9 \times 10^{-3}$.
The full distributions of both the $p$-values and $Z$ statistics are shown in Figure \ref{Fig:two-betas}.
These numbers are comparable to those reported in \dea, and they show that even a very small amount of misspecification in the null model could lead to very small $p$-values in the hypothesis testing framework.

\clearpage
\section*{Discussion}
Species rely on a variety of pathways for secondary antibody diversification, and the reasons for this variety remain an immunological puzzle.
The current understanding is that chickens, rabbits, and some other species use a combination of gene conversion and somatic hypermutation during affinity maturation, while humans and mice use only somatic hypermutation.
A recent paper by \dea\ suggests that humans and mice also use extensive templated mutagenesis to diversify their repertoires, which may happen by a mechanism similar to gene conversion.
This finding was based on a novel method, PolyMotifFinder, for identifying templated mutagenesis via microhomology, and in this article we studied its properties.

We were interested in the false positive rate of the PolyMotifFinder strategy and developed a novel way of estimating this rate.
We ran the algorithm on two sets of mutation observations, derived from mouse and human respectively, using two corresponding sets of simulated donor genes not present in the subject in question; any inferences of templated mutagenesis in this case must be spurious.
The homology structure of these ``mock'' donor genes mimicked that of the set of potential templated mutagenesis donors present in the subject.
Using this method, we found that although the PolyMotifFinder strategy is quite sensitive to templated mutagenesis, it also has a false positive rate exceeding 50\% for the donor tract sizes considered in \dea.
We used our estimates of the false positive rates of the PolyMotifFinder strategy along with the naive PolyMotifFinder estimates of the rate of templated mutagenesis to obtain upper bounds on the true rate in mice and humans.
In each case, we obtain upper bounds ranging from zero to around 10\%, although because these are upper bounds, the true rate may also be zero.

Many of the results in \dea\ were based on findings of a statistically significant deviation from a null model instead of an estimate of the rate of templated mutagenesis.
The results of the PolyMotifFinder/RandomCheck strategy were presented in terms of a Stouffer's $Z$ score, which describes deviation from a simplified null hypothesis about the way the mutations arise.
We showed that the observed Stouffer's $Z$ values and $p$-values in \dea\ are not proof of templated mutagenesis, but merely reflect the fact that the specified null model is incorrect, and given thousands of samples we have enough power to detect even small departures from it.

The same considerations apply to the findings of linkage disequilibrium in the mutated sequences: a statistically significant amount of linkage disequilibrium does not imply templated mutagenesis, and is in fact entirely consistent with the Neuberger model.
In particular, if a mutation-generating process satisfies
\begin{itemize}
\item mutation at one site implies a higher probability of mutation at nearby sites, and
\item not every base has an equal probability of being chosen as the new base for mutation,
\end{itemize}
then sites that are close together will be in linkage disequilibrium, even though the mutations are not introduced by templated mutagenesis.
One of the potential pathways posited by the Neuberger model to resolve AID lesions has exactly the properties described above.
In that pathway, an exonuclease strips out several nucleotides around the AID-induced lesion, and the resulting single-stranded sequence is patched by Pol $\eta$, an error-prone polymerase.
Thus, a mutation at one position is likely to be accompanied by mutations at neighboring positions, since Pol $\eta$ might have introduced multiple errors in the same patch of nucleotides.
In addition, we do not expect Pol $\eta$ to replace nucleotides uniformly at random, since we expect bias in the nucleotide misincorporation rate \cite{Rogozin2001-yp}.
We accordingly expect this pathway to cause linkage disequilibrium between sites, particularly those that are close together.
Therefore, the observed significant linkage disequilibrium is not \emph{prima facie} evidence of templated mutagenesis.

Next, we describe several limitations of the analysis presented here to be considered when interpreting the results.
First of all, our bounds depend on our estimate of the false positive rate being an underestimate of the true false positive rate.
We have two main reasons for believing that this is true, particularly for the human sequences.
The first is that our mock donor sets of simulated {\em gpt} homologs are slightly smaller than the corresponding IMGT donor gene sets.
The mock {\em gpt} set based on the mouse IMGT IGHV genes has 462 unique genes, compared with 499 in the mouse IMGT IGHV gene set.
The corresponding numbers for the mock {\em gpt} set based on the human IMGT IGHV genes and the human IMGT IGHV genes are 404 and 466.
The mock gene sets have slightly smaller numbers of genes than the gene sets they were based on because of the simulation method: not all of the branches in the inferred tree actually lead to a mutation in the simulations, and so there are fewer unique genes than leaves in the tree.
Our second reason for believing our estimate of the false positive rate is conservative involves the correspondence between diversity in variable regions and mutation hotspots: in real antibody sequences, mutations are more likely to occur in the CDRs, and there is also more variability in the IGHV genes in the CDRs.
This is not the case for the {\em gpt} sequences: as demonstrated in \cite{Yeap2015Sequence}, there are mutation hotspots in the {\em gpt} genes as well, but these hotspots do not correspond to regions of higher variability in the mock {\em gpt} gene sets.
Since mutations are more likely to occur in regions with more templated mutagenesis templates in the antibody gene sequences than in the {\em gpt} sequences, we believe that the false positive rate estimate based on the {\em gpt} sequences is lower than the true false positive rate.

We emphasize that we have obtained bounds on, not estimates of, the rate of templated mutagenesis, and that these bounds depend on assumptions about the sensitivity of PyMF and on our estimate of the false positive rate being conservative.
For humans, the quality of the bound also depends on how well our estimate of the PyMF false positive rate translates from mice to humans.
We were only able to estimate the false positive rate of PyMF in the mouse because of the transgenic system set up in \cite{Yeap2015Sequence}, and that estimate translates to humans to the extent that the somatic hypermutation processes of the two species coincide.
We expect the processes to be similar enough that the false positive rate is valid for both species, but any differences that do exist mean that the bounds for mice are more reliable than those for humans.

It is still possible that templated mutagenesis occurs at a low rate.
If so, characterizing its properties is important because, even if templated mutagenesis events occur infrequently, they could increase the rate of certain mutation patterns immensely.
This has important implications for estimation procedures (phylogenetic estimation, germline annotation, etc) as well as translational applications such as rational vaccine design.
Thus, we do not view our work as closing the book on the interesting possibility that templated mutagenesis could play a role in B cell diversification.

\clearpage
\section*{Acknowledgements}
The authors are grateful to Trevor Bedford, Christian Busse, Gordon Dale, and Joshy Jacob for discussions that improved this analysis, and to Leng-Siew Yeap and Duncan Ralph for help analyzing the \emph{gpt} data.
Christian Busse provided helpful comments on the manuscript.

\bibliographystyle{ji}
\bibliography{main}

\begin{thebibliography}{10}

\bibitem{Methot2017-gi}
Methot, S.~P. and J.~M. Di~Noia. 2017. Chapter two - molecular mechanisms of
  somatic hypermutation and class switch recombination. In {\em Advances in
  Immunology}, {Frederick W. Alt}, ed. 133: 37--87. Academic Press.

\bibitem{Dale2019-db}
Dale, G.~A., D.~J. Wilkins, C.~D. Bohannon, D.~Dilernia, E.~Hunter, T.~Bedford,
  R.~Antia, I.~Sanz, and J.~Jacob. 2019. Clustered mutations at the murine and
  human {IgH} locus exhibit significant linkage consistent with templated
  mutagenesis. {\em J. Immunol.} 203: 1252--1264.

\bibitem{Bornholdt2016-ii}
Bornholdt, Z.~A., H.~L. Turner, C.~D. Murin, W.~Li, D.~Sok, C.~A. Souders,
  A.~E. Piper, A.~Goff, J.~D. Shamblin, S.~E. Wollen, T.~R. Sprague, M.~L.
  Fusco, K.~B.~J. Pommert, L.~A. Cavacini, H.~L. Smith, M.~Klempner, K.~A.
  Reimann, E.~Krauland, T.~U. Gerngross, D.~K. Wittrup, E.~O. Saphire, D.~R.
  Burton, P.~J. Glass, A.~B. Ward, and L.~M. Walker. 2016. Isolation of potent
  neutralizing antibodies from a survivor of the 2014 {Ebola} virus outbreak.
  {\em Science}. 351: 1078-1083.

\bibitem{Yeap2015Sequence}
Yeap, L.-S., J.~K. Hwang, Z.~Du, R.~M. Meyers, F.-L. Meng, A.~Jakubauskaite,
  M.~Liu, V.~Mani, D.~Neuberg, T.~B. Kepler, J.~H. Wang, and F.~W. Alt. 2015.
  {Sequence-Intrinsic} mechanisms that target {AID} mutational outcomes on
  antibody genes. {\em Cell}. 163: 1124-1137.

\bibitem{Boettiger2015-vi}
Boettiger, C. 2015. An introduction to {Docker} for reproducible research. {\em
  Oper. Syst. Rev.} 49: 71--79.

\bibitem{Vander_Heiden2014-pd}
Vander~Heiden, J.~A., G.~Yaari, M.~Uduman, J.~N.~H. Stern, K.~C. O'Connor,
  D.~A. Hafler, F.~Vigneault, and S.~H. Kleinstein. 2014. {pRESTO}: a toolkit
  for processing high-throughput sequencing raw reads of lymphocyte receptor
  repertoires. {\em Bioinformatics}. 30: 1930--1932.

\bibitem{Retter2007-zr}
Retter, I., C.~Chevillard, M.~Scharfe, A.~Conrad, M.~Hafner, T.-H. Im,
  M.~Ludewig, G.~Nordsiek, S.~Severitt, S.~Thies, A.~Mauhar, H.~Bl{\"o}cker,
  W.~M{\"u}ller, and R.~Riblet. 2007. Sequence and characterization of the ig
  heavy chain constant and partial variable region of the mouse strain {129S1}.
  {\em J. Immunol.} 179: 2419--2427.

\bibitem{Edgar2004-rk}
Edgar, R.~C. 2004. {MUSCLE}: multiple sequence alignment with high accuracy and
  high throughput. {\em Nucleic Acids Res.} 32: 1792--1797.

\bibitem{Price2010-cy}
Price, M.~N., P.~S. Dehal, and A.~P. Arkin. 2010. {FastTree} 2--approximately
  maximum-likelihood trees for large alignments. {\em PLoS One}. 5: e9490.

\bibitem{Spielman2015-gl}
Spielman, S.~J. and C.~O. Wilke. 2015. Pyvolve: A flexible python module for
  simulating sequences along phylogenies. {\em PLoS One}. 10: e0139047.

\bibitem{Goldman1994-to}
Goldman, N. and Z.~Yang. 1994. A codon-based model of nucleotide substitution
  for protein-coding {DNA} sequences. {\em Mol. Biol. Evol.} 11: 725--736.

\bibitem{Ralph2016-hj}
Ralph, D.~K. and F.~A. Matsen, 4th. 2016. Consistency of {VDJ} rearrangement
  and substitution parameters enables accurate {B} cell receptor sequence
  annotation. {\em PLoS Comput. Biol.} 12: e1004409.

\bibitem{Wang2008-ic}
Wang, Y., K.~J.~L. Jackson, W.~A. Sewell, and A.~M. Collins. 2008. Many human
  immunoglobulin heavy-chain {IGHV} gene polymorphisms have been reported in
  error. {\em Immunol. Cell Biol.} 86: 111--115.

\bibitem{Bates-lme4}
Bates, D., M.~M{\"a}chler, B.~Bolker, and S.~Walker. 2015. Fitting linear
  mixed-effects models using {lme4}. {\em Journal of Statistical Software}. 67:
  1--48.

\bibitem{Rcore-2017}
{R Core Team}. {\em R: A Language and Environment for Statistical Computing}.
\newblock R Foundation for Statistical Computing. Vienna, Austria. 2017.

\bibitem{Rogozin2001-yp}
Rogozin, I.~B., Y.~I. Pavlov, K.~Bebenek, T.~Matsuda, and T.~A. Kunkel. 2001.
  Somatic mutation hotspots correlate with {DNA} polymerase eta error spectrum.
  {\em Nat. Immunol.} 2: 530--536.

\end{thebibliography}

\clearpage

\section*{Supplementary Materials}

\renewcommand{\thetable}{\Roman{table}}
\renewcommand{\tablename}{Supplemental Table}
\renewcommand{\figurename}{Supplemental Figure}

\makeatletter
\def\tagform@#1{\maketag@@@{[\ignorespaces#1\unskip\@@italiccorr]}}
\makeatother

\renewcommand{\baselinestretch}{1.5}

\setcounter{figure}{0}
\setcounter{table}{0}

\begin{table}[h!] \centering
  \begin{tabular}{l | l l l l l l}
Donor set &     Min.  & 1st Qu. &  Median    & Mean & 3rd Qu. &    Max. \\\hline
{\em gpt} mock from Human &  0.18 &  0.30 &  0.38 &  0.41 &  0.50 &  0.83 \\
IMGT Human & 0.23 &  0.31 &  0.36 &  0.40 &  0.46 &  0.74 \\
{\em gpt} mock from mouse&  0.18 &  0.39 &  0.49 &  0.48 &  0.58 &  0.79 \\
IMGT Mouse & 0.21 &  0.38 &  0.50 &  0.48 &  0.57 &  0.74
  \end{tabular}
  \caption{Five-number summaries of the set of divergences between genes and root for four donor gene sets.
  The divergences for the {\em gpt} human mock set are similar to the divergences for the IMGT human set, and the divergences for the {\em gpt} mouse mock set are similar to the divergences for the IMGT mouse set.}
\end{table}

\newpage

\begin{figure}[h!]
\begin{center}
\includegraphics[width=.8\textwidth]{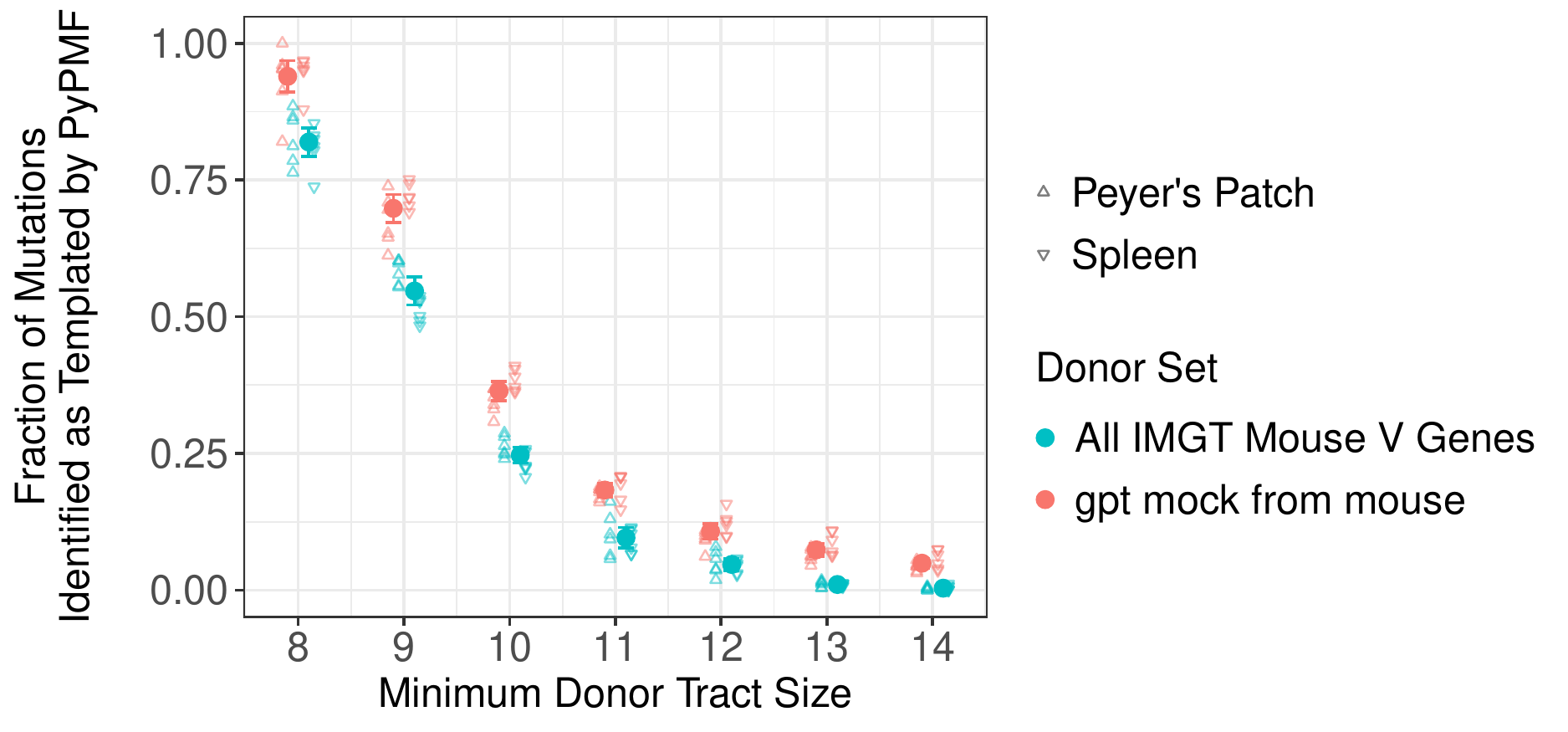}
\end{center}
\caption{%
  Hollow triangles represent the fraction of mutations explainable by templated mutagenesis in each sample, with upward-pointing triangles corresponding to samples from Peyer's patches and downward-pointing triangles corresponding to samples from the spleen.
  Reverse complements are included in each donor set.
For each tract length, the filled circle and error bar represents the overall estimate of the probability of a mutation being explainable by templated mutagenesis plus or minus two standard errors.
Points corresponding to samples from Peyer's patches and spleen are offset slightly to the left and right, respectively, to facilitate comparison and to avoid overplotting.
 This analysis was performed once on data from six individual mice, with two replicates per mouse corresponding to samples from Peyer's patches and spleen, yielding 12 total samples.%
}
\label{Fig:fpr-rc}
\end{figure}

\newpage

\begin{figure}[h!]
\begin{center}
\includegraphics[width=.8\textwidth]{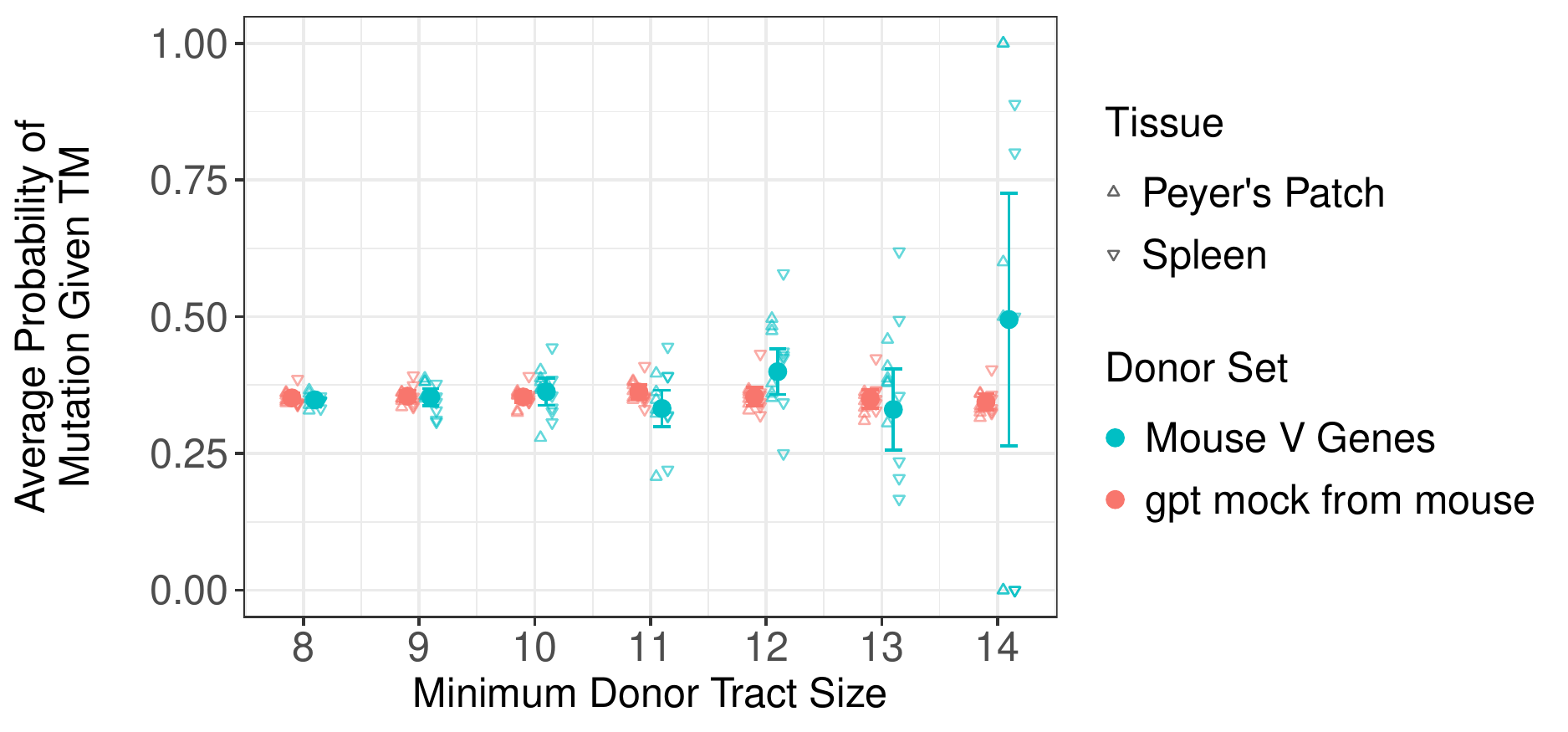}
\end{center}
\caption{%
Average probability of the observed mutations under a templated mutagenesis model, using {\em gpt} genes and their reverse complements (red) as well as the set of 129S1 V genes and their reverse complements (blue).
Each point corresponds to one sample taken from either spleen or Peyer's patches.
This analysis was performed once on data from six individual mice, with two replicates per mouse corresponding to samples from Peyer's patches and spleen, yielding 12 total samples.%
}
\label{Fig:prob-given-gcv-rc}
\end{figure}

\newpage

\begin{table}[h!] \centering
  Mice:\\
  \begin{tabular}{l | l l l l l l}
$k$&PyPMF rate&PyPMF FPR&UB (1)&UB (.99)&UB (.95)&UB (.9)\\\hline
8&0.9&0.94&0&0&0&---\\
9&0.7&0.7&0.02&0.02&0.03&0.03\\
10&0.46&0.36&0.15&0.15&0.16&0.18\\
11&0.22&0.18&0.04&0.04&0.04&0.04\\
12&0.13&0.11&0.03&0.03&0.03&0.03\\
13&0.08&0.07&0.01&0.01&0.01&0.01\\
14&0.06&0.05&0.01&0.01&0.01&0.01
\end{tabular}
\vspace{2em}

Humans:\\
\begin{tabular}{l | l l l l l l}
$k$&PyPMF rate&PyPMF FPR&UB (1)&UB (.99)&UB (.95)&UB (.9)\\\hline
8&0.88&0.91&0&0&0&---\\
9&0.6&0.65&0&0&0&0\\
10&0.34&0.29&0.07&0.07&0.08&0.08\\
11&0.22&0.14&0.09&0.09&0.1&0.1\\
12&0.16&0.08&0.09&0.09&0.1&0.1\\
13&0.15&0.05&0.1&0.1&0.11&0.11\\
14&0.14&0.03&0.1&0.11&0.11&0.12
\end{tabular}
\caption{Upper bounds (UB) on the rate of templated mutagenesis in the VB1-8 (top) and the anti-Ebola sequences (bottom) computed for a range of tract lengths $k$ and sensitivities when including reverse complements in the donor set.
 $k$ denotes tract length, PyPolyMF rate is the naive PyPolyMF estimate of the rate of templated mutagenesis, PyPolyMF FPR is the PyPolyMF false positive rate, UB denotes upper bound, and the number in paretheses denotes the assumed sensitivity (true positive rate) of PyPolyMF.}
  \label{Tab:boundRc}
\end{table}

\end{document}